\documentclass[%
 reprint,
 superscriptaddress,
 showpacs,
 amsmath,amssymb,
 aps,
 prl,
]{revtex4-1}

\usepackage{graphicx} 
\usepackage{dcolumn}  
\usepackage{bm}       

\begin{document}

\title{Measurement of double-polarization asymmetries in the quasi-elastic
$^3\vec{\mathrm{He}}(\vec{\mathrm{e}},\mathrm{e}'\mathrm{p})$ process}

\author{M.~Mihovilovi\v{c}} \affiliation{Faculty of Mathematics and Physics, University of Ljubljana, SI-1000 Ljubljana, Slovenia} \affiliation{Jo\v{z}ef Stefan Institute, SI-1000 Ljubljana, Slovenia} \affiliation{Institut f\"ur Kernphysik, Johannes-Gu\-ten\-berg-Universit\"at, Mainz, Germany}
\author{G.~Jin} \affiliation{University of Virginia, Charlottesville, VA 22908, USA}
\author{E.~Long} \affiliation{University of New Hampshire, Durham, NH 03824, USA}
\author{Y.-W.~Zhang} \affiliation{Rutgers University, New Brunswick, NJ 08901, USA}
\author{K.~Allada} \affiliation{Thomas Jefferson National Accelerator Facility, Newport News, VA 23606, USA}
\author{B.~Anderson} \affiliation{Kent State University, Kent, OH 44242, USA}
\author{J.~R.~M.~Annand} \affiliation{Glasgow University, Glasgow G12 8QQ, Scotland, United Kingdom}
\author{T.~Averett} \affiliation{The College of William and Mary, Williamsburg, VA 23187, USA}
\author{W.~Bertozzi} \affiliation{Massachusetts Institute of Technology, Cambridge, MA 02139, USA}
\author{W.~Boeglin} \affiliation{Florida International University, Miami, FL 33181, USA}
% ...
\collaboration{The Jefferson Lab Hall A Collaboration} \noaffiliation

\date{\today}

\begin{abstract}
We report on a precise measurement of double-polarization asymmetries 
in electron-induced breakup of $^3\mathrm{He}$ proceeding to $\mathrm{pd}$ 
and $\mathrm{ppn}$ final states, performed in quasi-elastic kinematics 
at $Q^2 = 0.25\,(\mathrm{GeV}/c)^2$ for missing momenta up to 
$250\,\mathrm{MeV}/c$.  These observables represent highly sensitive tools 
to investigate the electromagnetic and spin structure of $^3\mathrm{He}$ 
and the relative importance of two- and three-body effects involved 
in the breakup reaction dynamics.  The measured asymmetries cannot be 
satisfactorily reproduced by state-of-the-art calculations 
of $^3\mathrm{He}$ unless their three-body segment is adjusted, 
indicating that the spin-dependent part of the nuclear interaction 
governing the three-body breakup process is much smaller than previously 
thought.
\end{abstract}


\maketitle

The $^3\mathrm{He}$ nucleus represents the key challenge of nuclear physics 
due to its potential to reveal the basic features of nuclear structure
and dynamics in general.  In particular, this paradigmatic three-body 
system offers a unique opportunity to study the interplay of two-nucleon
and three-nucleon interactions, an effort at the forefront of nuclear physics 
research \cite{Glock04,Golak2005,FBS}. Modern theoretical descriptions 
of the structure and dynamics of $^3\mathrm{He}$ require, first of all,
a detailed understanding of the nuclear Hamiltonian (including 
the three-nucleon force), which generates the consistent nuclear ground
and scattering states, while accounting for final-state 
interactions (FSI).  The reaction mechanism comprises also 
the electromagnetic current operator, which takes into account 
meson-exchange currents (MEC).  Experiments on $^3\mathrm{He}$, 
particularly those involving polarization degrees of freedom, 
provide essential input to theories which need to be perpetually 
improved to yield better understanding of the underlying physics 
and to match the current increase in experimental precision. 
The quality of theoretical models is crucial to all $^3\mathrm{He}$-based 
experiments seeking to extract neutron information by utilizing 
$^3\mathrm{He}$ as an effective neutron target, an approximation 
relying on a sufficient understanding
of the proton and neutron polarization within polarized $^3\mathrm{He}$.

The $^3\mathrm{He}$ nucleus is best studied by electron-induced 
knockout of protons, deuterons and neutrons, where the sensitivity 
to various aspects of the process can be greatly enhanced by 
the use of polarized beam and target \cite{Golak2005}.  
The focus of this paper is on the two-body (2bbu) and three-body (3bbu)
breakup channels with proton detection in the final state,
$^3\vec{\mathrm{He}}(\vec{\mathrm{e}},\mathrm{e}'\mathrm{p})\mathrm{d}$ and
$^3\vec{\mathrm{He}}(\vec{\mathrm{e}},\mathrm{e}'\mathrm{p})\mathrm{pn}$,
which were investigated concurrently with the already published
$^3\vec{\mathrm{He}}(\vec{\mathrm{e}},\mathrm{e}'\mathrm{d})$ data 
\cite{Miha14}.

In a $^3\vec{\mathrm{He}}(\vec{\mathrm{e}},\mathrm{e}'\mathrm{p})$ 
reaction the virtual photon emitted by the incoming electron transfers 
the energy $\omega$ and momentum $\bm{q}$ to the $^3\mathrm{He}$ 
nucleus.  The process observables are then analyzed in terms of 
missing momentum, defined as the difference between the momentum transfer 
and the detected proton momentum,
$p_\mathrm{m} = |\bm{q} - \bm{p}_\mathrm{p}|$, thus
$p_\mathrm{m}$ corresponds to the momentum of the recoiled 
deuteron in 2bbu and the total momentum of the residual $\mathrm{pn}$
system in 3bbu.

The unpolarized $^3\mathrm{He}(\mathrm{e},\mathrm{e}'\mathrm{p})$
process at low energies has been studied at MAMI, both on 
the quasi-elastic peak \cite{Florizone99} and below it \cite{Kozlov04}.  
The bulk of our present high-energy information comes from 
two experiments in quasi-elastic kinematics at Jefferson Lab 
\cite{Marat05,Fatiha05}, resulting in reaction cross-sections 
at high $p_\mathrm{m}$ and yielding important insight into nucleon 
momentum distributions, isospin structure of the transition currents, 
FSI, and MEC.  However, just as in 
the $(\mathrm{e},\mathrm{e}'\mathrm{d})$ case, experiments 
that exploit polarization offer much greater sensitivity to 
the fine details of these ingredients.  Such measurements have
been extremely scarce.  A single asymmetry data point with high
uncertainty exists from NIKHEF \cite{Poolman,Higinbotham:2000hc}.
In addition, we have a precise measurement of both transverse 
and longitudinal asymmetries separately for the 2bbu and 3bbu channels
in quasi-elastic kinematics \cite{Carasco,Achenbach}, but the measurement 
was restricted to (and summed over) relatively low $p_\mathrm{m}$.

Early theoretical studies \cite{Laget,Nagorny,NagornyS} have shown strong 
sensitivities of double-polarization asymmetries in $^3\mathrm{He}$ 
breakup to the isospin structure of the electromagnetic current,
to the sub-leading components of the $^3\mathrm{He}$ ground-state 
wave-function, as well as to the tensor component of the nucleon-nucleon
interaction.  However, while in the deuteron channel these would
predominantly manifest themselves at low $p_\mathrm{m}$, the 2bbu 
and 3bbu proton channels should give more information at high $p_\mathrm{m}$, 
a region which is, however, difficult to explore experimentally.
These diagrammatic evaluations ultimately gave way to more refined, 
full Faddeev calculations performed independently by the Krakow 
\cite{GolakPRC02,GolakPRC05} and the Hannover/Lisbon 
\cite{Yuan02a,Deltuva04a,Deltuva04b,Deltuva05} groups, 
which we use in this paper.
The key feature of our experiment is the unmatched precision 
of the extracted asymmetries together with a broad kinematic range,
with $p_\mathrm{m}$ extending to as far as $250\,\mathrm{MeV}/c$.
This extended coverage represents a crucial advantage, since Faddeev 
calculations indicate that the manifestations of various wave-function 
components, as well as the potential effects of three-nucleon forces, 
imply very different signatures as functions of $p_\mathrm{m}$.


If both beam and target are fully polarized, the cross-section for
the $^3\vec{\mathrm{He}}(\vec{\mathrm{e}},\mathrm{e}'\mathrm{p})$
reaction has the form
$$
\frac{\mathrm{d}\sigma(h,\vec{S})}{\mathrm{d}\Omega}
  = \frac{\mathrm{d}\sigma_0}{\mathrm{d}\Omega}
    \left[\,1+\vec{S}\cdot\vec{A}^0+h(A_\mathrm{e}+\vec{S}\cdot\vec{A})\,
    \right] \>, 
$$
where $\mathrm{d}\Omega = \mathrm{d}\Omega_\mathrm{e}\mathrm{d}E_\mathrm{e}
\mathrm{d}\Omega_\mathrm{p}$ is the differential of the phase-space
volume, $\sigma_0$ is the unpolarized cross section, 
$\vec{S}$ is the spin of the target, and $h$ is the helicity 
of the electrons.  Here $\vec{A}^0$ and $A_\mathrm{e}$ are 
the target and beam analyzing powers, respectively, while 
the spin-correlation parameters $\vec{A}$ yield 
the asymmetries when both the beam and the target are polarized. 
If the target is polarized only in the horizontal plane defined
by the beam and scattered electron momenta,
the term $\vec{S}\cdot\vec{A}^0$ does not contribute \cite{Laget},
while $A_\mathrm{e}$ is suppressed and is negligible
with respect to $\vec{A}$.

The orientation of the target polarization is defined by the angles
$\theta^\ast$ and $\phi^\ast$ in the frame where the $z$-axis is
along $\bm{q}$ and the $y$-axis is given by
$\bm{p}_\mathrm{e}\times\bm{p}_\mathrm{e}'$.
The measured asymmetry at given $\theta^\ast$ and $\phi^\ast$ is then 
$$
A(\theta^\ast,\phi^\ast) =
\vec{S}(\theta^\ast,\phi^\ast)\cdot\vec{A} =
 \frac{(\mathrm{d}\sigma/\mathrm{d}\Omega)_+ -
       (\mathrm{d}\sigma/\mathrm{d}\Omega)_-}
      {(\mathrm{d}\sigma/\mathrm{d}\Omega)_+ + 
       (\mathrm{d}\sigma/\mathrm{d}\Omega)_-} \>,
$$
where the subscript signs represent the beam helicities.  
In this paper we report on the measurements of these asymmetries in
$^3\vec{\mathrm{He}}(\vec{\mathrm{e}},\mathrm{e}'\mathrm{p})\mathrm{d}$ and
$^3\vec{\mathrm{He}}(\vec{\mathrm{e}},\mathrm{e}'\mathrm{p})\mathrm{pn}$
processes.  The measurements were performed
during the E05-102 experiment at the Thomas Jefferson National Accelerator 
Facility in experimental Hall~A \cite{HallANIM}, with a beam energy 
of $2.425\,\mathrm{GeV}$ in quasi-elastic kinematics at four-momentum 
transfer of $Q^2 = \bm{q}^2 - \omega^2 = 0.25\,(\mathrm{GeV}/c)^2$.


The beam was longitudinally polarized, with an average polarization 
of $P_\mathrm{e} = (84.3\pm 2.0)\,\%$ measured by a M{\o}ller polarimeter
\cite{HallANIM}. The target was a $40\,\mathrm{cm}$-long glass cell 
containing $^3\mathrm{He}$ gas at approximately $9.3\,\mathrm{bar}$
($0.043\,\mathrm{g/cm}^2$), polarized by hybrid spin-exchange optical 
pumping \cite{Walker,Appelt,Babcock,Singh}.  Two pairs of Helmholtz coils
were used to maintain the in-plane target polarization direction at
$67^\circ$ and $156^\circ$ with respect to $\bm{q}$, allowing us to
measure $A(67^\circ,0^\circ)$ and $A(156^\circ,0^\circ)$, respectively.  
Electron paramagnetic and nuclear magnetic resonance 
\cite{Abragam,Romalis,Babcock2} were used to monitor the target 
polarization, $P_\mathrm{t}$, which was between $50\,\%$ and $60\,\%$
when corrected for dilution due to nitrogen.


The scattered electrons were detected by a High-Resolution magnetic 
Spectrometer (HRS) \cite{HallANIM}, while the protons were detected 
by the large-acceptance spectrometer BigBite equipped with a detector 
package optimized for hadron detection \cite{BBNIM}.  Details
of the experimental setup and the procedure to extract the very pure 
sample of electron-proton coincidence events are given in Ref.~\cite{Miha14}.


The experimental asymmetry for each orientation of the target polarization
was determined as the relative difference between the number of 
background-subtracted coincidence events corresponding to positive 
and negative beam helicities,
$A_\mathrm{exp} = (N_+ - N_-) / (N_+ + N_-)$, where $N_+$ and $N_-$
have been corrected for helicity-gated beam charge asymmetry, 
dead time and radiative effects.  The corresponding final values 
of the physics asymmetries were calculated 
as $A = A_\mathrm{exp} / (P_\mathrm{e}P_\mathrm{t})$.

\begin{figure}[hbtp]
\begin{center}
\includegraphics[width=8.5cm]{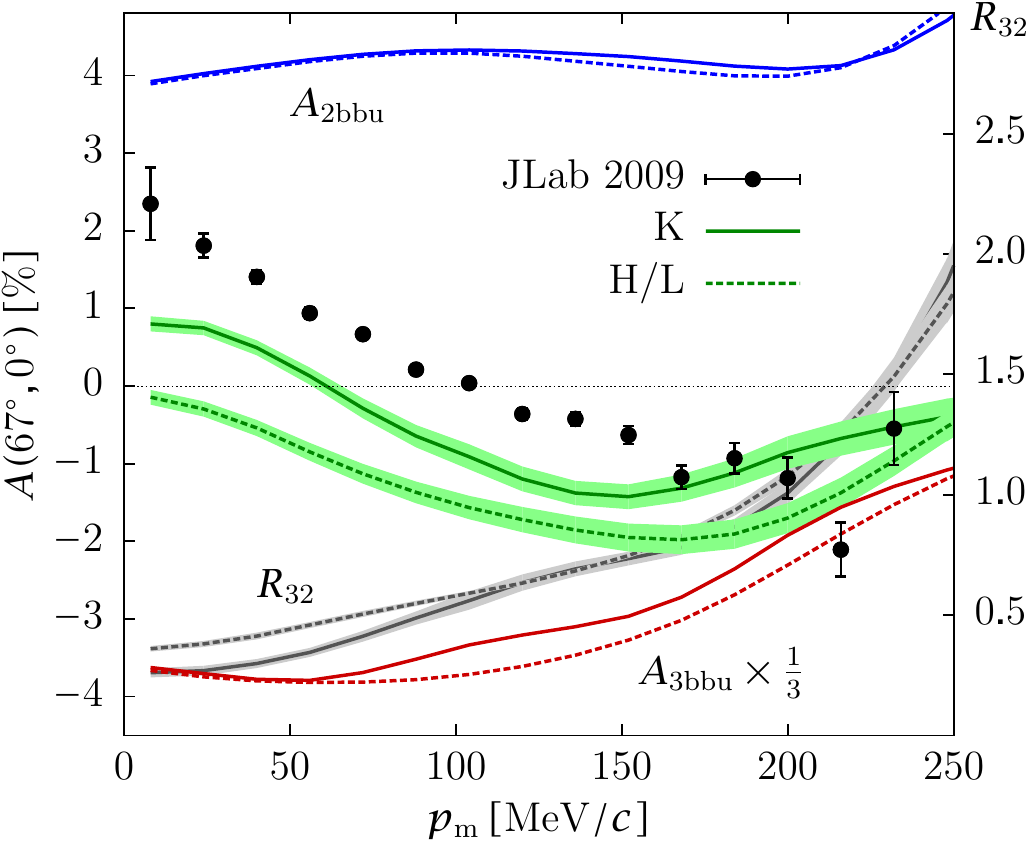}

\bigskip

\includegraphics[width=8.5cm]{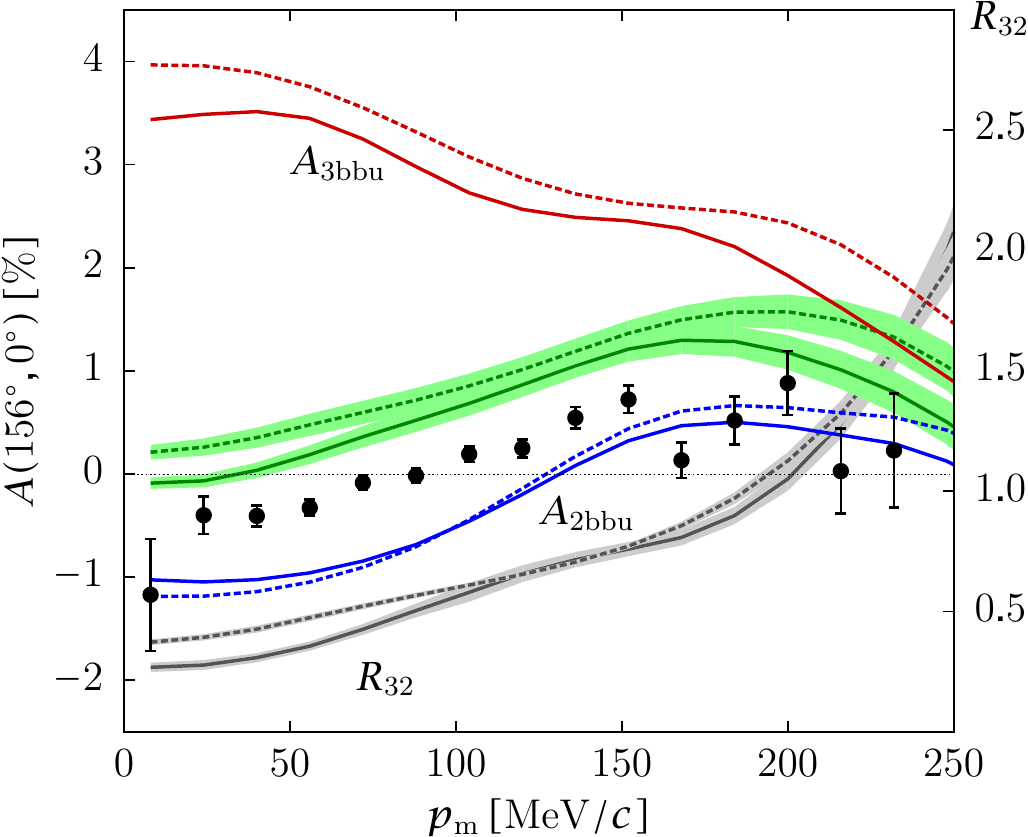}
\end{center}
\vspace*{-5mm}
\caption{(Color online.) The asymmetries $A(67^\circ,0^\circ)$ (top)
and $A(156^\circ,0^\circ)$ (bottom) in the quasi-elastic
$^3\vec{\mathrm{He}}(\vec{\mathrm{e}},\mathrm{e}'\mathrm{p})$ process
(2bbu and 3bbu combined) as functions of missing momentum, compared 
to theoretical predictions (green) showing the 2bbu (blue) and 3bbu (red) 
contributions as well as the ratio of 3bbu and 2bbu cross-sections (grey,
right axis).  All full (dashed) lines correspond to K (H/L) calculations, 
respectively.}
\label{altpm}
\end{figure}

The resulting asymmetries as functions of $p_\mathrm{m}$ are shown 
in Fig.~\ref{altpm}.  
The largest contribution to their systematic error comes from 
the relative uncertainty in the target polarization, $P_\mathrm{t}$, 
which has been estimated at $\pm 5\,\%$, followed by the uncertainty 
in the target dilution factor ($\pm 2\,\%$) and the absolute uncertainty 
of the beam polarization, $P_\mathrm{e}$ ($\pm 2\,\%$).  
The uncertainty in the target orientation angle represents 
a minor contribution ($\pm 0.6\,\%$) to the total uncertainty, 
totaling $\approx 6\,\%$ (relative).

Figure~\ref{altpm} also shows the results of the state-of-the-art 
three-body calculations of the Krakow (K) \cite{GolakPRC02,GolakPRC05}
and Hannover/Lisbon (H/L) \cite{Yuan02a,Deltuva04a,Deltuva04b,Deltuva05} 
groups.  The K calculations are based on the AV18 nucleon-nucleon
potential \cite{av18} and involve a complete treatment of FSI 
and the dominant part of MEC, but do not include three-nucleon forces; 
the Coulomb interaction is taken into account in the $^3\mathrm{He}$ 
bound state.  The H/L calculations are based on 
the coupled-channel extension of the charge-dependent Bonn potential 
\cite{cdbonn} and also include FSI and MEC, while the $\Delta$ isobar 
is added as an active degree of freedom providing a mechanism 
for an effective three-nucleon force and for exchange currents.  
Point Coulomb interaction is added in the partial waves involving 
two charged baryons.  In contrast to the K and H/L approaches, 
the Pisa (P) calculations \cite{marcucci05} are not genuine 
Faddeev calculations but are of equivalent precision and are expected 
to account for all relevant reaction mechanisms.  The P calculations 
are based on the AV18 
interaction model (augmented by the Urbana IX three-nucleon force
\cite{uix}), in which full inclusion of FSI is taken into account by means 
of the variational pair-correlated hyper-spherical harmonic expansion, 
as well as MEC.  At present, the Pisa group only provides 2bbu calculations.  
Coulomb interaction is included only in the bound state in K calculations,
but in both bound and scattering states in H/L and P calculations.
Due to the extended experimental acceptance, all theoretical asymmetries 
were appropriately averaged, resulting in the error bands around
the theoretical curves in Fig.~\ref{altpm}. Details can be found 
in \cite{Miha14}.


Neither the K nor the H/L calculation reproduces the measured
asymmetries to a satisfactory level.  Similarly to our findings 
in the deuteron channel, the theories approximately capture their 
overall functional forms, but exhibit systematic vertical offsets 
of up to two percent.  In calculations a strong cancellation 
is involved in obtaining each total asymmetry from its 2bbu and 3bbu 
contributions, which are typically opposite in sign and 
of different magnitudes.  Nevertheless, the failure of the theories
to reproduce the data can be traced to the 3bbu asymmetry alone,
as discussed in the following.

\begin{figure}[hbtp]
\begin{center}
\includegraphics[width=8.5cm]{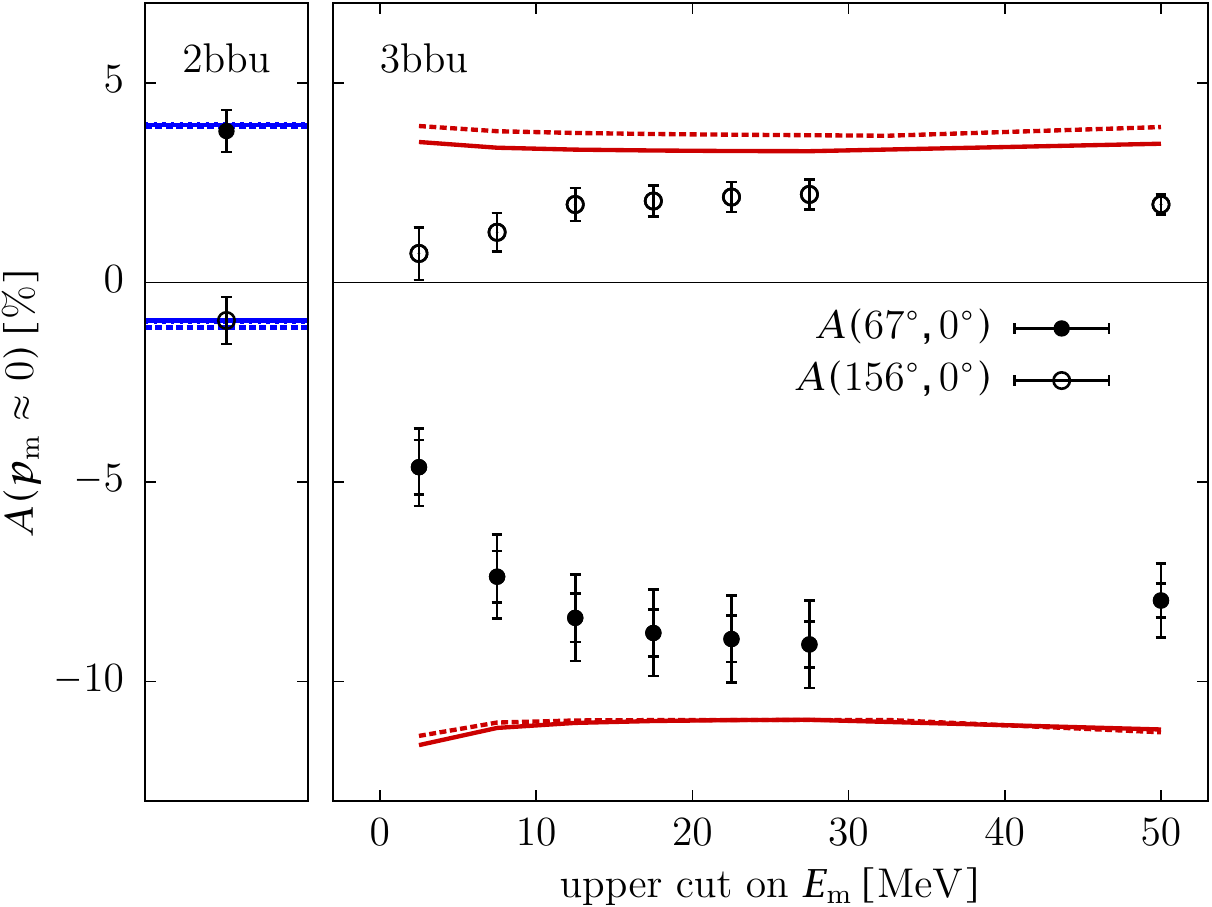}
\end{center}
\vspace*{-5mm}
\caption{(Color online.) The extracted asymmetries for 2bbu (left)
and 3bbu (right).  Curve notation as in Fig.~\ref{altpm}, with
the addition of the Pisa 2bbu calculation in the left panel
(blue dotted lines hidden beneath the full and dashed lines).}
\label{a23}
\end{figure}

Since the energy resolution of our measurement was insufficient 
to directly disentangle the 2bbu and 3bbu channels, the individual 
asymmetries were extracted by restricting the data sample 
to $p_\mathrm{m} \approx 0$ and studying the dependence 
of $A(67^\circ,0^\circ)$ and $A(156^\circ,0^\circ)$ 
in terms of the cut in missing energy, 
$E_\mathrm{m} = \omega - T_\mathrm{p} - 7.7\,\mathrm{MeV}$.
The comparison of the measured $E_\mathrm{m}$ spectrum with 
the simulated one revealed that in spite of the overlap between 
the two channels, the lowest portion of the distribution 
at $E_\mathrm{m} < 0$ is dominated by 2bbu, thus allowing for
the extraction of the corresponding asymmetry, $A_\mathrm{2bbu}$, 
which agrees with the calculations to better than $0.5\,\%$ 
(absolute): see Fig.~\ref{a23} (left).  According to 
the simulation the contribution of 3bbu to the experimental 
cross-section is approximately $7\,\%$, suggesting that 
near the threshold the size of the 3bbu asymmetry is 
about $1\,\mathrm{\%}$, much smaller than the prediction. 
However, a better insight into the 3bbu asymmetry has been obtained 
by investigating the data at $E_\mathrm{m} > 0$.  Considering that 
the measured asymmetries contain also the 2bbu contribution,
the 3bbu asymmetry (Fig.~\ref{a23} (right)) has been extracted 
from the data as
$$
A_\mathrm{3bbu} 
  = \frac{(1+R_{32}) A_\mathrm{exp}-A_\mathrm{2bbu}}{R_{32}} \>, 
$$
where $R_{32}$ is the 3bbu/2bbu cross-section ratio shown 
in Fig.~\ref{altpm} corrected for finite momentum and angular
resolutions as well as radiative effects.  Typically $R_{32}$ 
ranges from $0.20$ to $0.33$ and is assumed to be well under control 
in both K and H/L calculations.  The extracted asymmetries 
are in good agreement with the theory in the limit where 
the whole spectrum ($E_\mathrm{m} \leq 50\,\mathrm{MeV}$) 
is considered in the analysis, but strongly deviate from 
the theory near threshold ($E_\mathrm{m} \leq 2.5\,\mathrm{MeV}$) 
for the 3bbu reaction channel. 


\begin{figure}[hbtp]
\begin{center}
\includegraphics[width=8.5cm]{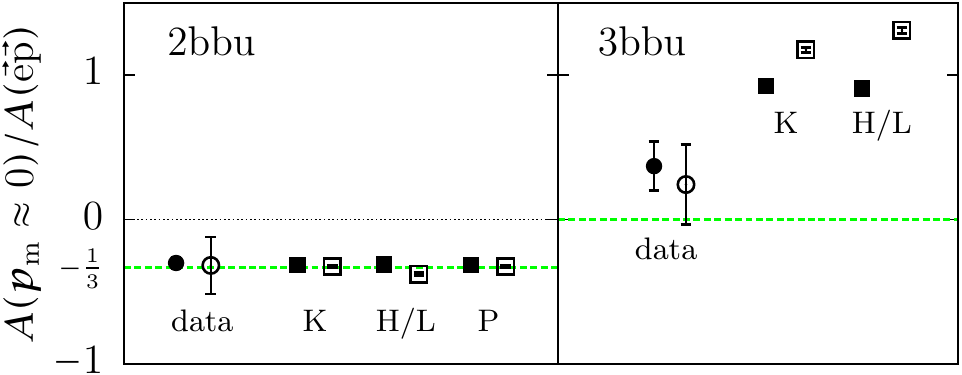}
\end{center}
\vspace*{-5mm}
\caption{(Color online.) The $A(67^\circ,0^\circ)$ (full symbols)
and $A(156^\circ,0^\circ)$ (empty symbols) asymmetries for 2bbu (left) 
and 3bbu (right) divided by the corresponding asymmetries for elastic 
$\vec{\mathrm{e}}\vec{\mathrm{p}}$ scattering at the same value of $Q^2$.
In both panels the data (circles) are compared to the calculations (squares).
The tiny uncertainties on the theoretical points are due to the averaging
procedure.}
\label{amessage}
\end{figure}

In an effort to compensate for the effect of spin orientation 
of protons inside the polarized $^3\mathrm{He}$ nucleus, we have 
divided the nuclear asymmetries by the asymmetries for elastic 
$\vec{\mathrm{e}}\vec{\mathrm{p}}$ scattering at the same value 
of four-momentum transfer; see Fig.~\ref{amessage}.  In a simplified 
picture of the $^3\vec{\mathrm{He}}(\vec{\mathrm{e}},\mathrm{e}'\mathrm{p})$
process, one would expect the 2bbu ratio at $p_\mathrm{m} \approx 0$
to be $-1/3$, corresponding to the effective polarization of the 
(almost free) proton inside the polarized $^3\mathrm{He}$ nucleus,
while the 3bbu ratio should vanish because any of the two oppositely
polarized protons could be knocked out in the process.  Indeed,
in the 2bbu case both the experimental and the predicted ratios
coincide almost perfectly, at the anticipated ``naive'' value of $-1/3$.
On the other hand, in the 3bbu case the predictions cluster 
approximately around unity (and apparently retain a residual 
dependence on $\theta^\ast$), while the two experimental ratios 
are much smaller (and mutually consistent).

In conclusion, we have provided the world-first, high-precision
measurement of double-polarization asymmetries for proton knockout
from polarized $^3\mathrm{He}$ nuclei at two different spin settings 
and over a broad range of momenta.  Two state-of-the-art theoretical
approaches to the $^3\mathrm{He}$ disintegration process 
are able to approximately accommodate the main structural features 
of our data set.  Since the asymmetries are rather small and strong
cancellations of the two-body and three-body breakup contributions
are involved, the agreement can be deemed satisfactory and
the theoretical framework justified.  However, the high precision 
of our measurements has been able to reveal a substantial deficiency
in the calculations of the three-body breakup process, pointing
to a mismatch between our true relativistic kinematics 
and non-relativistic spin-dependent nuclear dynamics employed 
in the calculations.

Since the three-body breakup process is more selective than the corresponding
two-body breakup of $^3\mathrm{He}$, it will be very interesting
to verify if an application of consistent chiral two-nucleon
and three-nucleon interactions with chiral two-nucleon
and three-nucleon contributions in the electromagnetic
current operator will provide a solution of this problem.

\begin{acknowledgments}
We thank the Jefferson Lab Hall A and Accelerator Operations technical 
staff for their outstanding support.  This work was supported in part 
by the National Science Foundation and the U.S. Department of Energy. 
Jefferson Science Associates, LLC, operates Jefferson Lab for the U.S. 
DOE under U.S. DOE contract DE-AC05-06OR23177.  This work was supported 
in part by the Slovenian Research Agency (research core funding 
No.~P1--0102).  This work is a part of the LENPIC project and 
was supported by the Polish National Science Centre under Grants
No.  2016/22/M/ST2/00173 and 2016/21/D/ST2/01120. The numerical
calculations of the Krakow group were partially performed 
on the supercomputer cluster of the JSC, J\"ulich, Germany.
\end{acknowledgments}

\bibliographystyle{apsrev4-1}
\bibliography{he3eep}

\end{document}